\documentclass[prc,aps,nofootinbib,showkeys,showpacs,twocolumn]{revtex4}
\usepackage{epsfig}
\usepackage{graphicx}
\usepackage{amssymb}
\usepackage{color}
\begin{document}

\title{Exact description of non-Markovian effect in open quantum system: \\
the discretized environment method}

\author{Denis Lacroix} \email{lacroix@ipno.in2p3.fr}
\affiliation{Institut de Physique Nucl\'eaire, IN2P3-CNRS, Universit\'e Paris-Sud, F-91406 Orsay Cedex, France}
\author{V. V. Sargsyan}
\affiliation{International Center for Advanced Studies, Yerevan State University, M. Manougian 1, 0025 Yerevan, Armenia}
\affiliation{Joint Institute for Nuclear Research, 141980 Dubna, Russia}
\author{G. G. Adamian }
\affiliation{Joint Institute for Nuclear Research, 141980 Dubna, Russia}
\author{N. V. Antonenko}
 \affiliation{Joint Institute for Nuclear Research, 141980 Dubna, Russia}

\date{\today}
\begin{abstract}
An approach, called discretized environment method, is introduced to treat exactly  non-Markovian
effects in open quantum systems. In this approach, a complex environment described
by a spectral function is mapped into a finite set of discretized states with an appropriate coupling to the system of interest.
The finite set of system  plus environment degrees of freedom
are then explicitly followed in time leading to a quasi-exact description. The present approach is
anticipated to be particularly accurate in the low temperature and strongly non-Markovian regime. 
The discretized environment method  is validated on a two-level 
system (qubit) coupled to a bosonic or fermionic heat bath. A perfect agreement with the quantum
 Langevin approach 
is found. Further illustrations are made on a three-level  system (qutrit) coupled to a bosonic heat-bath.
Emerging processes due to strong memory
effects are discussed.
\end{abstract}



\maketitle

\section{Introduction}
In the last decades, efforts have been made to describe the dynamics of open quantum systems
in various regime of coupling to an environment at zero or finite temperature \cite{Wei99,Bre02}.
From these efforts, a large variety of methods either exact or approximate have emerged to treat the
non-Markovian effects which often play important role in the dynamics of 
open quantum system.  Among the deterministic approaches, one can quote projection techniques like
Nakajima-Zwangig (NZ) or Time-Convolutionless (TCL) \cite{Nak58,Zwa60,Has77,Shi77}. While being rather
efficient in some cases \cite{Bre99,Bre01}, these approaches rapidly becomes cumbersome as the complexity of
the system increases.

Recently, several stochastic methods have been proposed to reformulate exactly the problem of a system
coupled to an environment. This includes the Quantum state diffusion (QSD) approach \cite{Str96,Dio98,Str99,Str05,Pii08},
the quantum Langevin approach  \cite{M11,M12,M13,M14,M15}, 
the Quantum Monte-Carlo (QMC) \cite{Sha04,Lac05,Lac08, Hup10} or the stochastic method of Ref. \cite{Bre04}.
While very promising, these approaches have at present been applied to rather specific situations where the system
remains quite simple. The first reason is that, stochastic equation of motion can not always be obtained easily.
This is the case of the QSD method where even a three-level  system appeared until recently as a challenge \cite{Jin10}.
A second reason, is that even if stochastic Schr\"odinger equations   can be derived without
difficulties like in the QMC approach \cite{Lac05,Lac08},  
some trajectories might be
unstable and can lead to uncontrolled numeric. 

The aim of the present work is to provide a simple deterministic approach which is able to describe  exactly
for a large class of situations in the dynamics of an open quantum system.
The main hypothesis of the method is that the environment can be explicitly incorporated
in a discretized form.  For this reason, we named the approach "the Discretized Environment Method" (DEM). The
DEM  is anticipated to be rather versatile and effective especially in the   non-Markovian
regime. In Sect. II the DEM is introduced. We show that the hypothesis of a discretized environment can incorporate
thermalized systems. The link with standard method dedicated to open quantum system is made through the introduction of the spectral
function. The approach is then validated in 2-level  (Sect. III.A)
and 3-level   (Sect. III.B) systems coupled to a heat bath.

\section{Discretized Environment Method for open quantum systems}

In the present work, we consider the situation where a system S is linearly coupled
to an environment E or thermal bath. The Hamiltonian $H$ is decomposed as:
\begin{eqnarray}
H & = & \sum_i \hbar \omega_i S^\dagger_i S_i + \sum_\nu \hbar \omega_\nu B^\dagger_\nu B_\nu + H_C. \label{eq:hamref}
\end{eqnarray}
The creation/annihilation operators $\{ S^\dagger_i , S_i \}$ (resp. $\{ B^\dagger_\nu , B_\nu \}$) are associated to the system (resp. environment)
degrees of freedom. $H_C$ denotes the coupling Hamiltonian between the two subsystems and is written here as:
\begin{eqnarray}
H_C = \sum_{i\nu} g_{i\nu} (S_i   \otimes B^\dagger_\nu+ S^\dagger_i \otimes B_\nu).  \label{eq:coupling}
\end{eqnarray}
 In the following, we simply assume that the coupling constant are independent on the system state, i.e. $g_{i\nu} = g_\nu$.

The difficulty in treating the system+environment problem stems from the complexity of the environment. It is usually
assumed that the number of degrees of freedom  of the environment is too large to follow them explicitly in time.
To deal with this complexity, the standard strategy  is to derive the equation of motion for the system degrees of freedom only where the effect
of the environment is treated in an approximative way \cite{Wei99,Bre02}.

\subsection{Direct diagonalization approach}

In the present article, we start from a completely different strategy and assume that, in some cases, the environment degrees of freedom
can be explicitly included in the description and that the Hamiltonian can be directly diagonalized. Such diagonalization requires a priori to introduce a basis for the full Hilbert space of the system plus environment. In the following, we introduce system and environment ground states, denoted by
$ | 0 \rangle_S $ and
$| 0 \rangle_E $.
These states can be considered as particle vacuum respectively for the creation operators $S^\dagger_i$ and $B^\dagger_\nu$:
\begin{eqnarray}
| i \rangle & = & S^\dagger_i  | 0\rangle_S, ~~~S_i  | 0 \rangle_S = 0, \nonumber \\
| \nu \rangle  &=&  B^\dagger_\nu  | 0 \rangle_E, ~~~S_\nu  | 0 \rangle_E = 0.
\end{eqnarray}
The set of product states $\{ | 0 \rangle_S \otimes | 0 \rangle_E ;| i \rangle \otimes | 0 \rangle_E \} ;$
$| 0 \rangle_S \otimes | \nu \rangle ; | i \rangle \otimes | \nu \rangle \}$ forms a
basis of the complete Hilbert space.

Two difficulties generally prevents a direct diagonalization of the Hamiltonian. The first one  comes from
the fact that the Hilbert space, associated to the environment, is usually of infinite size. The second one  is that even if one can consider a finite
but very large set of environment states, in many situations, the basis will remains rather large and diagonalization will require large numerical effort.
Denoting by $N_S$ ($N_E$) the number of excited states in the system (in the environment). The total Hilbert space
size is $(N_S+1)\times(N_E+1)$.


The specific shape of the Hamiltonian  considered here is anticipated to reduce the effort necessary to perform a direct diagonalization.
Indeed, only the states $\{ | i \rangle \otimes | 0 \rangle_E \}$ and $\{ | 0 \rangle_S \otimes | \nu \rangle \}$ are coupled
to each other. Therefore, only matrices of dimension
$(N_S+ N_E)$ need to be diagonalized, reducing significantly the numerical challenge. For compactness, in the following, we will introduce the notation:
\begin{eqnarray}
| 0 \rangle & = & | 0 \rangle_S \otimes  | 0 \rangle_E , \nonumber \\
| i \rangle_0 & = &  | i \rangle \otimes | 0 \rangle_E ,  \nonumber \\
| \nu \rangle_0 & = &  | \nu \rangle \otimes | 0 \rangle_S . \nonumber
\end{eqnarray}
The state $| 0 \rangle$ is a vacuum for the total system. In the reduced space, formed by the states $\{ | i \rangle_0, | \nu \rangle_0\} $, the Hamiltonian matrix, denoted by $H_R$, is written as:
\begin{eqnarray}
H_R & = & \sum_{i} \hbar \omega_i |i \rangle_0 ~{}_0\langle i | + \sum_\nu \hbar \omega_\nu   |\nu \rangle_0 ~ {}_0\langle \nu|
\nonumber \\
&+& \sum_{i,\nu} g_\nu \left\{ |i \rangle_0 ~{}_0\langle \nu | + |\nu \rangle_0 ~{}_0\langle i | \right\}.
\end{eqnarray}

Formally, the diagonalization of the Hamiltonian $H_R$ is equivalent
to define a new subset of operators $\{ A^\dagger_\alpha, A_\alpha \}$ in such a way that the Hamiltonian
takes the form:
\begin{eqnarray}
H & = & \sum_\alpha \hbar \Omega_\alpha A^\dagger_\alpha A_\alpha.
\end{eqnarray}
The new operators are rather specific since they combines degrees of freedom of both the system and the environment.
Denoting by $U$ the associated unitary canonical transformation, we use the following convention 
 \begin{eqnarray}
 \left\{
 \begin{array}{ccc }
A^\dagger_\alpha  = & \sum_i  U_{i\alpha} [ S^\dagger_i \otimes 1_E] + \sum_\nu U_{\nu \alpha} [ 1_S \otimes B^\dagger_\nu ]  \\
\\
A_\alpha      =&    \sum_i  U^*_{i\alpha} [ S_i \otimes 1_E] + \sum_\nu U^*_{\nu \alpha} [ 1_S \otimes B_\nu ]
\end{array}
\right.  \label{eq:transf}  ,
\end{eqnarray}
and associate to them new vacuum $| 0_\alpha \rangle$ defined through the property $A_\alpha | 0_\alpha \rangle =0$.
For compacity of notation, we will write $A^\dagger_\alpha = \sum_i  U_{i\alpha} S^\dagger_i  + \sum_\nu U_{\nu \alpha}  B^\dagger_\nu$. The inverse transformation to (\ref{eq:transf}) are given by
\begin{eqnarray}
S^\dagger_i &=& \sum_\alpha U^*_{i \alpha }  A^\dagger_\alpha , ~~~~B^\dagger_\nu = \sum_\alpha U^*_{\nu \alpha} A^\dagger_\alpha. \label{eq:transf1}
\end{eqnarray}
In the following, we will systematically use the notation $(i,j)$ for the system, $(\nu, \mu)$ for the environment and $(\alpha,\beta)$
for the new states that combines both subsystems.

Note that besides the Hamiltonian, the number of excitation $N$ in the total system, that
is a constant of motion, also takes a simple form:
\begin{eqnarray}
N &=& \sum_i S^\dagger_i S_i + \sum_\nu B^\dagger_\nu B_\nu  =  \sum_\alpha A^\dagger_\alpha A_\alpha.
\end{eqnarray}

\subsection{Information on  system evolution}

The introduction of new operators where the Hamiltonian takes a diagonal form gives
directly access to the dynamical evolution.
Using the Heisenberg picture, the evolutions of the new degrees of freedom simply read:
\begin{eqnarray}
A^\dagger_\alpha(t) & = & e^{i\Omega_\alpha t} A^\dagger_\alpha(0), ~~~~
A_\alpha (t) = e^{-i\Omega_\alpha t} A_\alpha(0) \nonumber
\end{eqnarray}
Using the unitary transformation between the original operators and  $\{ A^\dagger_\alpha\}$, one can
directly get
\begin{eqnarray}
S^\dagger_i(t) &=&
\sum_{\alpha k } U^*_{i \alpha } U_{j \alpha} e^{i\Omega_\alpha t} S^\dagger_j (0) +
\sum_\nu U^*_{i \alpha } U_{\nu \alpha} e^{i\Omega_\alpha t} B^\dagger_\nu (0) \nonumber \\
&\equiv& \sum_k M_{ij} (t)S^\dagger_j (0) +  \sum_\nu M_{i\nu} (t) B^\dagger_\nu (0) . \label{eq:optransf}
\end{eqnarray}
The $M$ matrix introduced here contains all the information
about the system or environment evolution.

Let us, for instance, assume that the total initial density ${\cal D}(0)$ is separated as:
\begin{eqnarray}
{\cal D}(0) & = &  D_S(0) \otimes {\cal D}_E(0)
\end{eqnarray}
where ${\cal D}_{S}$ (${\cal D}_E$) are the initial system (environment) density matrix with ${\rm Tr}({\cal D}_{S/E})$. Note that more
general total density could be used if necessary.

We further assume that initially
\begin{eqnarray}
\langle S^\dagger_i (0)  S_j (0) \rangle & = & {\rm Tr}(S^\dagger_i  S_j  {\cal D}(0) ) = {\rm Tr}(S^\dagger_i  S_j  {\cal D}_S(0) ) = \delta_{ij} n^0_i \nonumber
\end{eqnarray}
and similarly
\begin{eqnarray}
\langle B^\dagger_\nu (0)  B_\mu (0) \rangle = \delta_{\nu \mu} n^0_{\nu}.
\end{eqnarray}

  Starting from Eq. (\ref{eq:optransf}), we directly get the compact expression for the two-time correlation function
\begin{eqnarray}
\langle S^\dagger_i (t) S_j(t') \rangle &=& \sum_k M^*_{jk}(t')  n^0_k M_{ik}(t)  \nonumber \\
&+& \sum_\nu M^*_{j\nu}(t')  n^0_\nu M_{i\nu }(t) \label{eq:timecorrelation}
\end{eqnarray}
that is often computed in open quantum systems to probe the memory effects.

\subsection{Discretization  of   environment}

The present approach relies on the possibility to have a discrete and finite environment.
In the theory of open quantum systems, a key quantity related to the non-Markovian effects is the
dissipative kernel $K(t)$, that we define here as 
\begin{eqnarray}
K(t) & = & \sum_\nu \frac{g^2_\nu}{\hbar^2 \omega_\nu} e^{-i\omega_\nu t}. \label{eq:kernel}
\end{eqnarray}
Most often  this expression is replaced by the integral such that:
\begin{eqnarray}
 \sum_\nu \frac{g^2_\nu}{\hbar^2 \omega_\nu} e^{-i\omega_\nu t} & \rightarrow & \int_0^{+\infty} \frac{\rho(\omega) g^2(\omega)}
 {\hbar^2 \omega} e^{-i\omega t} d\omega , \label{eq:continuous}
\end{eqnarray}
where $\rho(\omega)$ is the environment density of state and $g(\omega)$ is the density-dependent coupling.  Note that here
it is  implicitly  assumed that the environment has only positive frequencies, i.e. $\omega_\nu \ge 0$.
In practice, a special
continuous function is assumed for $\rho(\omega) g^2(\omega) = J(\omega)$. A typical example is the Lorentz-Drude spectral function  
\begin{eqnarray}
J(\omega) &=&  \frac{g_0}{\pi}  \omega \frac{\gamma^2}{\gamma^2 + \omega^2}, \label{eq:lorentz}
\end{eqnarray}
that leads to an exponentially decaying dissipative kernel 
\begin{eqnarray}
K(t) & = & \frac{g_0}{2} \gamma e^{-\gamma t } + i \frac{g_0}{\pi} \gamma^2 \int_0^{+\infty}
d\omega \frac{\sin(\omega t)}{\gamma^2 + \omega^2}. \label{eq:kri}
\end{eqnarray}
Memory effect will then be important if $\gamma$ is small, while for $\gamma \rightarrow +\infty$ the Markovian limit is reached.

Most importantly, we see that the continuous limit (Eq. (\ref{eq:continuous})) is just the opposite situation as the one we want to consider, i.e.
an environment with a discretized spectra. Therefore, to use the technique based on direct diagonalization and to make connection
with more standard approaches we need to invert the scheme depicted in  Eq. (\ref{eq:continuous}).

 Let us consider a situation where the given spectral function $J(\omega)$ is introduced. 
 Because this only constrains the product $\rho(\omega) g^2(\omega)$, there exists
some flexibility in fixing $\rho(\omega)$ and $g(\omega)$. Here, we  assume that the environment frequencies are uniformly distributed between $0$
and a maximal frequency $\omega_{\rm max}$ with:
\begin{eqnarray}
\omega_\nu & = & \Delta \omega (n + 1/2), ~~n=0,\cdots,N_{\rm max}
\end{eqnarray}
With this convention, the spacing parameter $\Delta \omega = \omega_{\rm max}/(N_{\rm max}+1/2)$.

Once the density of states is supposed to be constant  the coupling parameter is $g_\nu$ are automatically fixed by imposing
the proper continuous limit. This leads to the discrete mapping of the coupling:
\begin{eqnarray}
g_\nu & = &  \hbar \sqrt { \Delta \omega J(\omega_\nu) }.
\end{eqnarray}

The present methodology, that consists in (i) discretizing the environment space and (ii)
performing a direct diagonalization of the discretized Hamiltonian, 
is the essence of the DEM.
In some simple situations, the discretized environment has been employed to obtain analytical expressions for the
evolution \cite{Mil83,Har93,Lai88}, but as far as we know it 
has not been directly applied as a direct numerical approach
to OQS.

In the limit of small spacing with infinitely small $\Delta \omega$  and infinitely large $\omega_{\rm max}$ the approach is exact.
Of course, in practice, only finite size matrices can be diagonalized requiring both non-zero spacing and a finite boundary for the
highest frequency.
The two parameters $(\Delta \omega, \omega_{\rm max})$ determine both the numerical accuracy and the numerical effort
of the approach. They should be carefully chosen to  describe properly  a physical process
and in such a way that the number of states in the environment $N_{\rm max}$ is minimized. More precisely,
$\omega_{\rm max}$ determines
the time resolution $\Delta \tau = 2 \pi/\omega_{\rm max}$ while $\Delta \omega$ defines the maximal time
$\tau_{\rm max} = 2 \pi/\Delta \omega$
over which the calculation can be considered as accurate.

As an important remark, the present approach should be greatly simplified if some physical cutoff exists
on the maximal energy for the states to be considered. Having in mind the Lorentz-Drude spectral function,
in physical situations  a system will interact with states in the vicinity of its typical energy range  determined
by $\gamma$. The lower $\gamma$ is, the smaller $\omega_{\rm max}$ could be taken.  
The Markovian limit will be reached if $\gamma$ is very large. Therefore, we anticipate that {\it the more the
dynamics is non-Markovian, the easier it will be to use the DEM approach.} It is at variance with other techniques
that are usually simplified in the Markovian limit.

Independently on the $\gamma$ value,  a second physical cutoff can be used that is figured out from
Eq. (\ref{eq:timecorrelation}).  We see indeed that a state in the environment has an effect on the system
only if its initial  population $n^0_\nu$ is non negligible.  If the environment is a thermal bath, the number of states
needs to be considered depends on the temperature $T$. In particular, if $\gamma \rightarrow +\infty$, only
this effect will allow to truncate the environment space in a reasonable way. So, the second conclusion is that the
DEM will be easier to apply at lower temperature.

\section{Applications}

\subsection{Two-level system coupled to  heat-bath}

As a proof of the method accuracy, we consider the special case where the system only contains one state associated
to $(S^\dagger_0 , S_0)$ and with excitation energy $\hbar \Omega$. The system is considered to be coupled with a heat-bath
at various temperatures.  This situation is actually similar to the case of a two-level  system that has been used in Ref. \cite{Gar97,Gar97a}
to derive the exact master equation including non-Markovian effects.
Note that, the system frequency is renormalized by a counter-term to avoid the
unwanted shift of the energy induced  by the coupling \cite{Bre02}. The system frequency is set to:
\begin{eqnarray}
\hbar \Omega' & = & \hbar \Omega + \sum_\nu \frac{g^2_\nu}{\hbar^2 \omega_\nu} .
\end{eqnarray}
Both bosonic and fermionic system+environment are
considered. The fermionic or bosonic nature only enters through the initial occupancies of
the bath through:
 \begin{eqnarray}
n^0_\nu(T) & = & \frac{1}{\exp\left(\frac{\hbar \omega_\nu}{k_B T} \right) + \varepsilon},
\end{eqnarray}
where $ \varepsilon =-1$ ($+1$) for bosons (for fermions). We assume that the spectral function is the Lorentz-Drude
function given by Eq. (\ref{eq:lorentz}). In the following, we will use $\Omega$, $\Omega^{-1}$, and $\hbar \Omega$ as the units,
respectively, for the frequencies, time, and energies.
\begin{figure}[htbp]
\includegraphics[width=9cm]{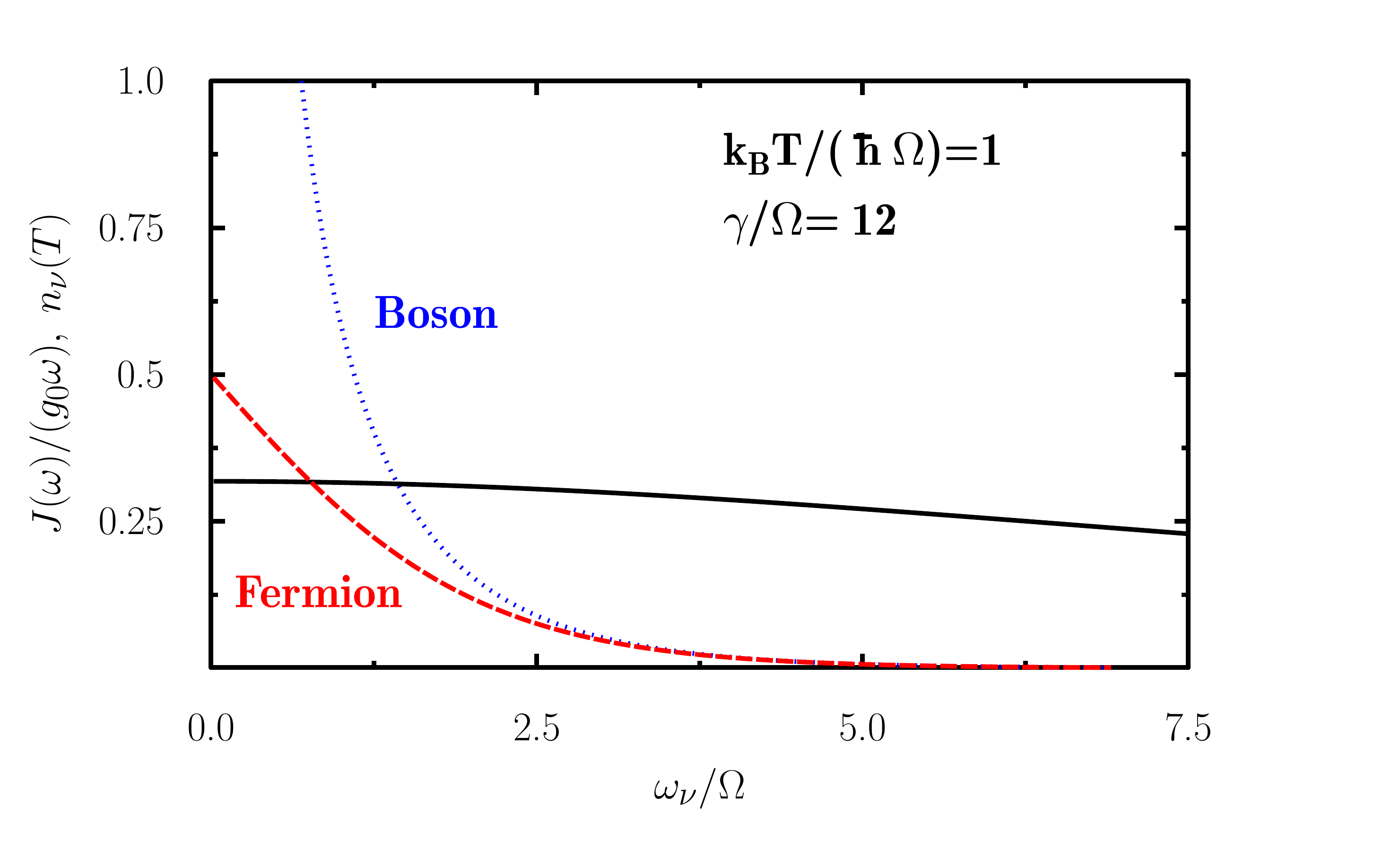}
\caption{ (color online)
Illustration of the cutoff induced by the temperature. 
The quantity displayed here are $J(\omega)/(g_0 \omega)$ (solid line), $n^0_\nu(T)$
for bosons (dotted line), and fermions (dashed line) at $\gamma/\Omega = 12$  and $k_B T/(\hbar \Omega)=1$. }
\label{fig:cutoff}
\end{figure}

Before presenting the results obtained with the DEM, it is interesting to illustrate the two contributions that will lead to
the natural cutoff in the bath frequencies. Assuming a typical physical situation where the state is coupled to a large set of
bath states $\gamma/\Omega = 12$  and moderate temperature $k_B T/(\hbar \Omega) = 1$, the occupation probabilities
for bosons or fermions as well as the function $J(\omega)/(g_0 \omega)$ are shown as a function of $\omega$ in Fig. \ref{fig:cutoff}.
We see that for not too high temperature, even if the spectral function permits the coupling to
high frequency states, these states are not necessarily needed to be considered due to the strong cut-off induced by their
initial occupancy.  In particular, even if $\gamma$ is rather large, a rather small cutoff in the bath frequencies can be used.
In practice, for $k_B T/(\hbar \Omega) \le 1$, a
cutoff frequency $\omega_{\rm max}/\Omega = 10$ and $250$ equidistant states insure a converged results.

\begin{widetext}
\begin{figure*}[htbp]
\includegraphics[width=8cm]{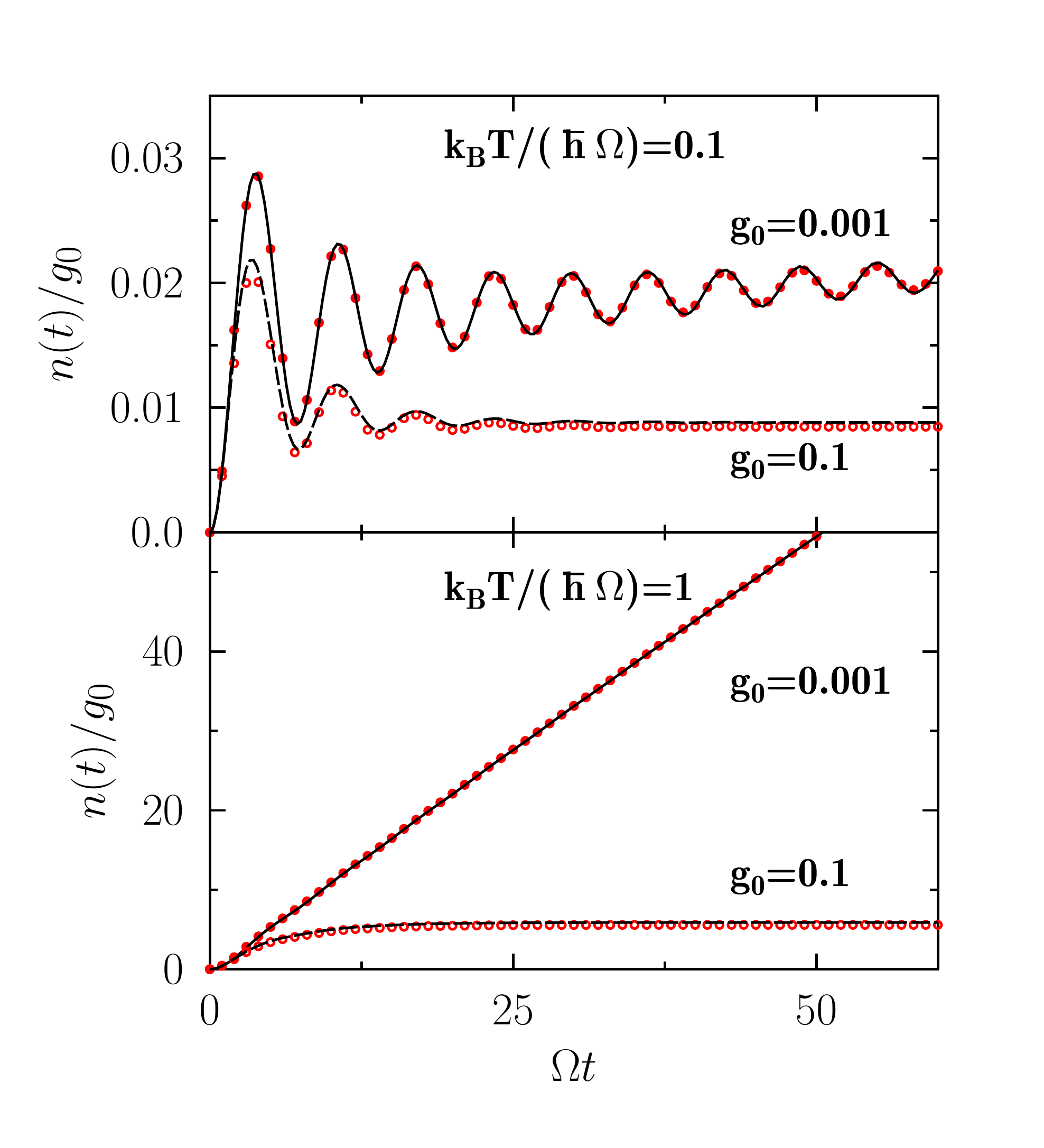}
\includegraphics[width=8cm]{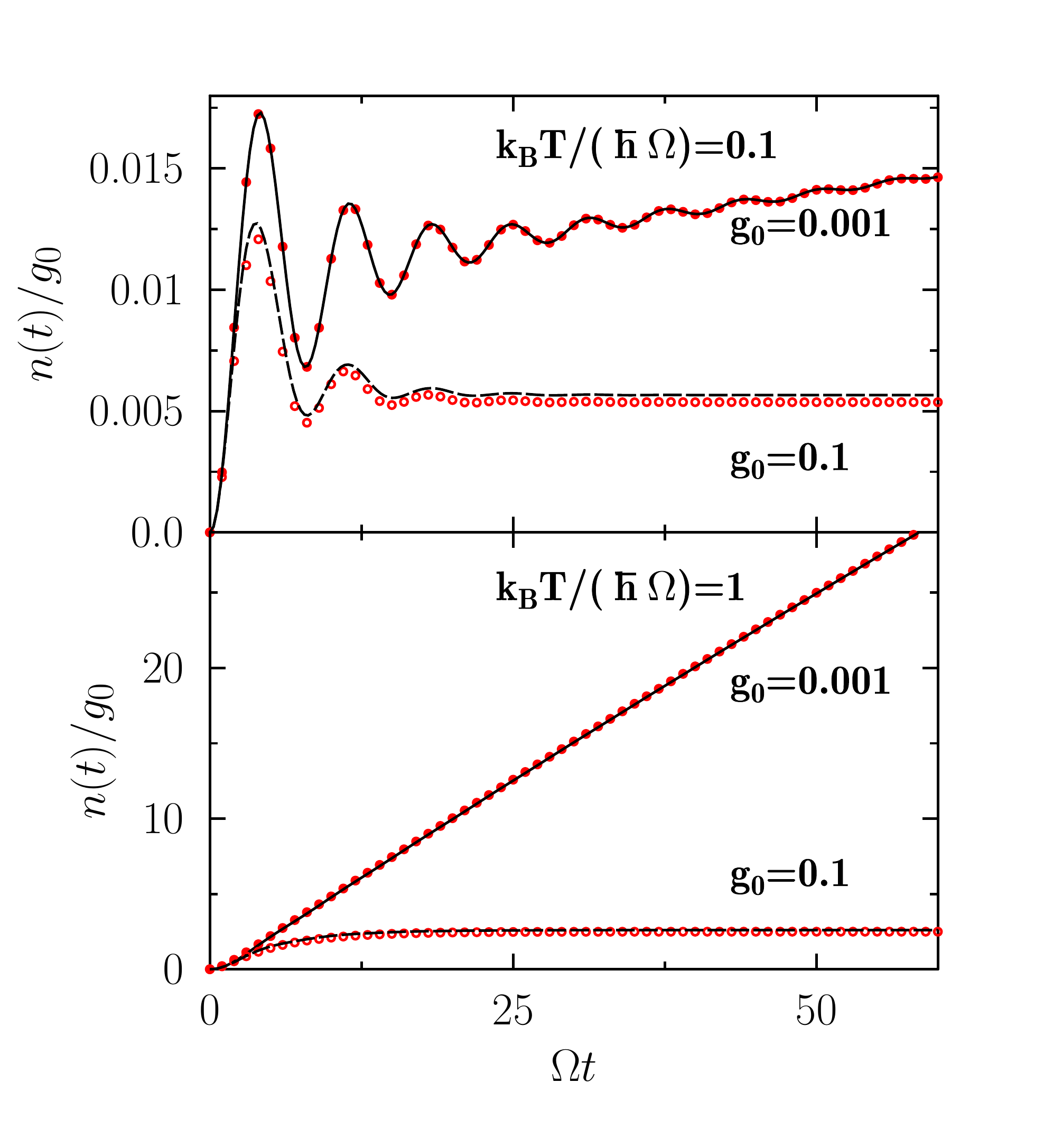}
\caption{ (color online)  Results obtained in bosonic (left side) and
 fermionic (right side) systems+bath  using the quantum Langevin
 approach (circles) are compared to results with the
DEM (lines) for different temperatures and coupling strengths. 
The calculations are performed for $\gamma/\Omega =12$.
}
\label{fig:oneboson}
\end{figure*}
\end{widetext}

In the following we compare the DEM results with the those obtained using a more standard approach based on the
quantum Langevin equation
where the environment dynamics is semi-analytically deduced using the Laplace transform technique \cite{Vaz14}.  The latter approach is exact under the
condition that the  imaginary part of the memory kernel $K(t)$ can be neglected. This assumption  is valid a priori for weak and
intermediate coupling $g_0$. In Fig.  \ref{fig:oneboson}, the two approaches are compared
either in the case of bosonic systems+bath (left) or fermionic systems+bath (right).
We see that in both cases, the two approaches agrees very well with each other.
The present agreement completely validates the DEM. In addition, the small number of states ($N_{\rm max} =250$) to be considered for the environment
shows that the present method can be numerically handled without any difficulty.



\subsection{Non-Markovian dynamics of  three-level  system}

As a second proof of the DEM feasibility, we consider a system formed by three levels coupled
to a heat bath. The three levels are denoted by $| L \rangle$, $| 0 \rangle$, and $| U \rangle$. The labels
"L", "0" and "U" stands here for lower, intermediate, and upper levels.   The three levels have, respectively, energies equal
to $-\hbar \Omega_L$, $0$, and $\hbar \Omega_U$. We further introduce the positive quantity $\hbar \Omega = \hbar (\Omega_U + \Omega_L)/2$.
In the following, as in Ref. \cite{Jin10} we assume $\hbar \Omega_U =\hbar \Omega_L= \hbar \Omega$.
Frequencies, energies and time will be given in terms of $\Omega$, $\hbar \Omega$ and $\Omega^{-1}$ units, respectively. This case
corresponds to the system considered in Ref. \cite{Jin10}, where the non-Markovian effects
have been investigated at zero temperature using the QSD. Here, we show that the DEM approach can be applied without
difficulty to the finite temperature case. We introduce the two creation operators:
\begin{eqnarray}
S^\dagger_L & = & | L \rangle \langle 0 |, ~~~ S^\dagger_U  =  | U \rangle \langle 0 | .\nonumber
\end{eqnarray}
The system+environment Hamiltonian is taken as:
\begin{eqnarray}
H & = & \hbar \Omega (S^\dagger_U S_U - S^\dagger_L S_L) +
\sum_\nu \hbar \omega_\nu B^\dagger_\nu B_\nu \nonumber \\
&+& \sum_\nu g_\nu (S_L + S_U) \otimes B^\dagger_\nu + {\rm H.c.}.
\end{eqnarray}

\subsubsection{Influence of  lower state}

Compared to the previous two-level  system case, the presence of an additional lower state can significantly enrich
the dynamics.  For instance, considering the zero temperature
case and a uniformly spaced environment with constant coupling $g_\nu = g$ and a spacing $\Delta \omega$,
when a single occupied
state is coupled to the environment, we expect a typical decay width $\Gamma \simeq 2 \pi g^2/\Delta \omega$, leading therefore to a reduction of the lifetime 
with increasing the interaction. Now, when two occupied levels are coupled to the same doorway states, the Dicke super-radiance effect can take place
and
leads to a reduction of the decay width (increase of the lifetime) as the coupling strength increases \cite{Dic54}.
Here, we consider a slightly different situation where the coupling is not uniform but, as previously, imposed by the spectral
function given in Eq. (\ref{eq:lorentz}). Since the bath  only has positive frequencies
$\hbar \omega > 0$, the symmetry between the lower and upper levels with respect to the ground state energy is broken
due to the coupling to the environment.

\begin{figure}[htbp]
\includegraphics[width=9cm]{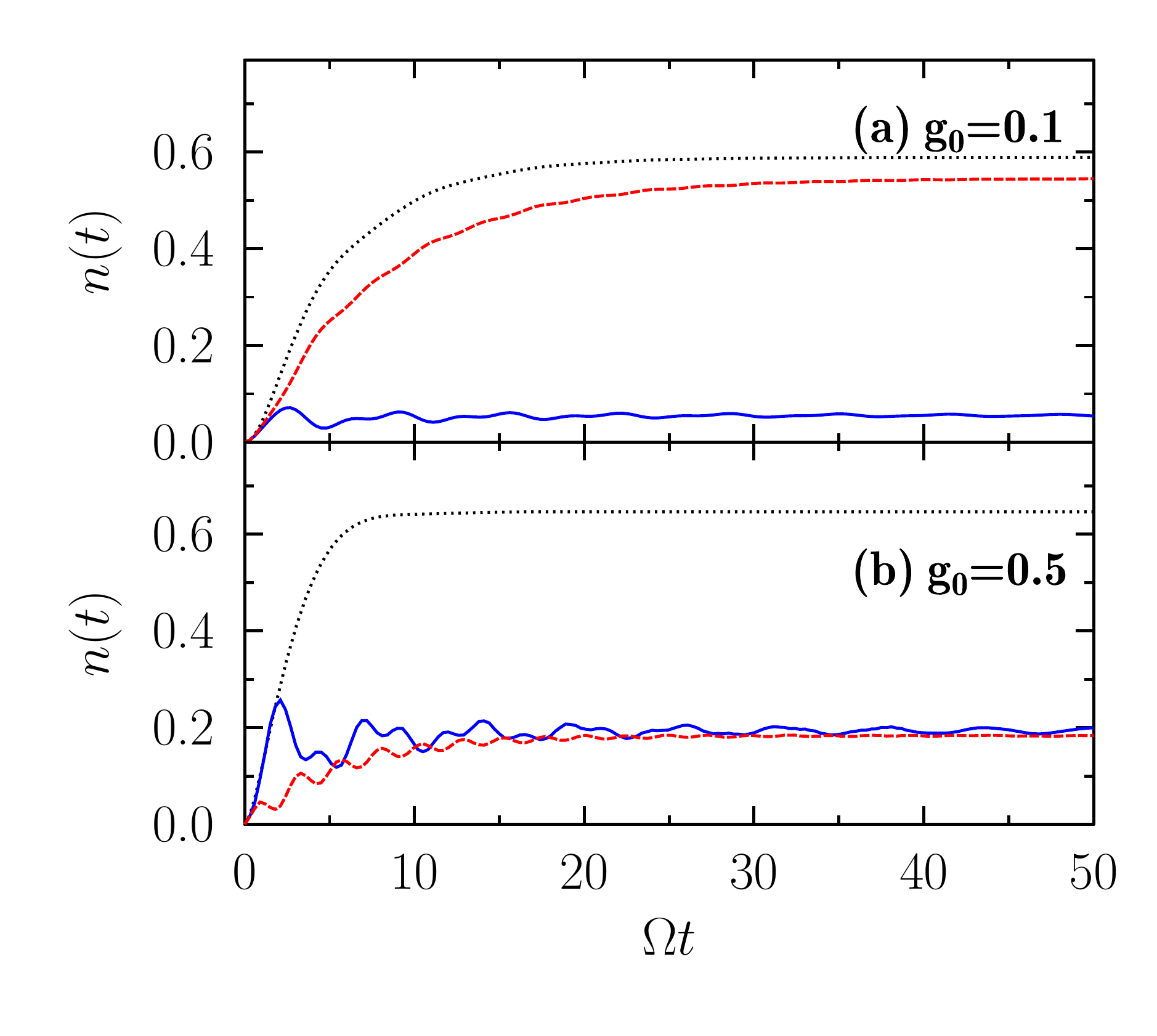}
\caption{ (color online)  Evolution of the upper (dashed line)  and lower  level (solid line) occupation
probabilities in the 3-level  system coupled to a  bosonic bath for the coupling strengths $g_0 = 0.1$ (a) and $0.5$ (b). 
 Both levels are assumed to be initially unoccupied, $n_U (t_0) = n_L(t_0) = 0$.
 In all cases, $k_B T/\hbar \Omega = 1.0$ and $\gamma/\Omega =12$.
 The 2-level system case presented
 in previous section where the excited state is initially unoccupied is also shown as a reference (dotted line).}
\label{fig:3levels}
\end{figure}

As an illustration of (i) the evolution of such 3-level  system,
(ii) the symmetry breaking, and (iii) the effect of the additional
lower state compared to the 2-level system case, the evolution
obtained with the DEM and different coupling strengths are 
shown in Fig. \ref{fig:3levels} at $k_B T/\hbar \Omega = 1.0$. 
In this figure,
both lower and upper states are assumed to be initially unoccupied.
In all cases, the evolution of the upper state is compared to the 2-level system case of previous section.
In the weak coupling regime $g_0=0.001$ (not shown), the dynamic of the upper level obtained in the 2- and 3-levels
systems coincide.  However, as can be seen from Fig.  \ref{fig:3levels} when $g_0$ increases, the presence of an additional
level at lower energy significantly affects the evolution. A marked reduction of the asymptotic
occupation of the upper level is observed. It is worth mentioning that in the 2-level  case, the asymptotic occupation
essentially reflects the statistical bosonic character of the bath that  imposes its temperature to the system (see Ref.
\cite{Vaz14}).   Here, significant deviation from this picture is seen. The asymptotic limit modification
is also accompanied by the appearance of coherent oscillations between the lower and upper level that could be understood
as an indirect coupling of the $| U \rangle$ and $| L \rangle$ levels through the heat-bath.

The  dynamics towards equilibrium is rather complex and no general rules can be figured out.
For instance, we see that at $g_0=0.1$ (panel (a) of Fig. \ref{fig:3levels}) for very short time
 the two levels evolves together and then separates
from each others towards different asymptotics. Comparing with the 2-level  case, we clearly see that the decay time  increases
due to the effect of the indirect coupling through the heat-bath, as could be expected from the Dicke super-radiance effects.
As $g_0$  increases (panel (b) of Fig. \ref{fig:3levels}), 
the transition time to equilibrium of the upper level   increases due to the extra oscillations that appeared. 
We see that the occupation of the lowest
level first follows the 2-level system case and then suddenly deviates from it.

\subsubsection{Three-level system dynamics in  strong non-Markovian limit}

In previous applications, the results have been obtained for rather large $\gamma/\Omega$ values where the memory
function, Eq. (\ref{eq:kri}), decays rapidly. As $\gamma$ decreases, the evolution is anticipated to be more and more
affected by the time-nonlocal nature of the system+environment evolution.  As noted above, smaller $\gamma$ values can be easily
described by the DEM method since it leads to a second natural cutoff in the oscillator frequencies to be considered.

Having in mind the picture  in Fig. \ref{fig:cutoff}, it is actually anticipated that the physics towards equilibrium
will encounter  a transition between two limits. At $(\hbar \gamma/k_B T) \gg 1$, that is the situations
considered previously, the number of doorway states is essentially restricted by initial thermal occupations.
When $(\hbar \gamma/k_B T) \ll 1$, the small width of the spectral function around zero frequency
will significantly cut the decay channel phase-space and largely influence the evolution. Empirically, we
have found that the decay properties are unchanged from $(\hbar \gamma/k_B T)=12$ down to
$(\hbar \gamma/k_B T) \simeq 2$ and starts to be modified for lower values of $\gamma$, as the system enters
the strongly non-Markovian regime.

\begin{figure}[htbp]
\includegraphics[width=9cm]{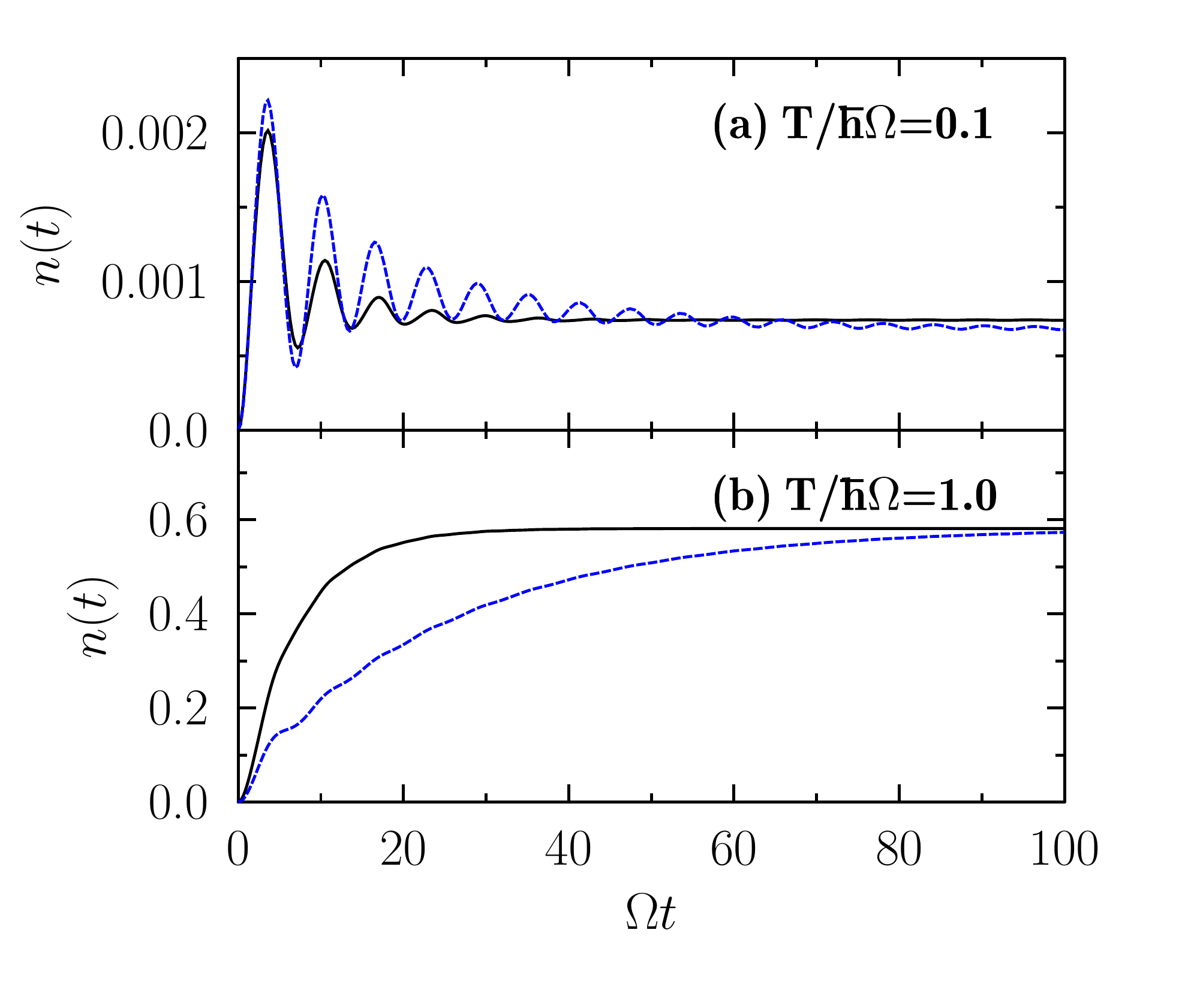}
\caption{ (color online)  Evolution of the upper level occupation probability in 
the 3-level  system coupled to a  bosonic bath for two different temperatures (a) $T/\hbar \Omega = 0.1$ and (b)
$T/\hbar \Omega = 1.0$. Both upper and lower levels are initially unoccupied, $n_U (t_0) = n_L(t_0) = 0$. 
In both panels, the different curves corresponds to different $\gamma/\Omega$ values:
0.5 (dashed line) and 2.0 (solid curve). 
In all cases, the coupling strength $g_0$ is set to 0.1.
}
\label{fig:ni3gam}
\end{figure}

\begin{figure}[htbp]
\includegraphics[width=9cm]{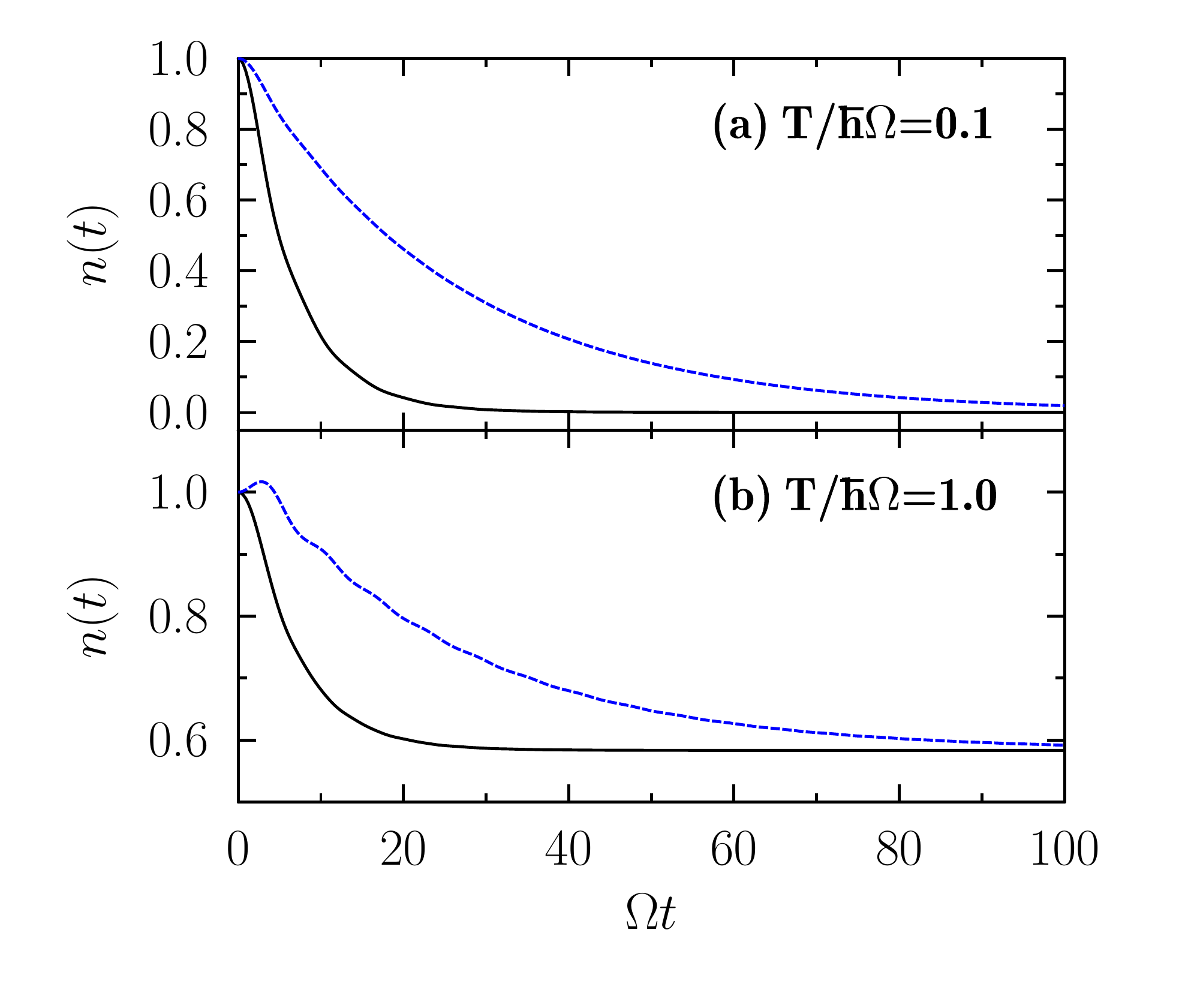}
\caption{ (color online) The same as figure \ref{fig:ni3gam}  except that the upper level is initially occupied
$n_U (t_0) = 1$  while $n_L(t_0) = 0$.
}
\label{fig:ni3gamN1}
\end{figure}

In the DEM, the total system is treated and one can formally take any $\gamma$.
In Figs. \ref{fig:ni3gam} and \ref{fig:ni3gamN1}, where the upper level is either assumed to be initially unoccupied or occupied while the
lower level is always unoccupied,  the effect of reducing   $\gamma$ value is illustrated for two different temperatures. In general,
it is seen that memory effects strongly influences the system dynamics. 
The transition time towards equilibrium   increases compared to the Markovian limit.


\section{Conclusion}

A direct approach is presented to treat a class of system+environment Hamiltonian exactly.
The approach, called Discretized Environment Method, relies on the possibility to accurately
discretize the environment into a limited set of  states able to mimic the complexity stemming from both the coupling
between the system and the environment on one side and the large density of state of the environment on the other
side. It is  anticipated that the approach is particularly accurate in the low temperature and strongly non-Markovian
limit.
The DEM is illustrated in the case of   2-level  system coupled to a bosonic or fermionic heat-bath and is compared
to the quantum Langevin approach  based on Laplace transform method where an exact solution  can be obtained analytically. A perfect agreement
is found between both approach, giving evidence that DEM can be a valuable tool to treat the open quantum systems.  A second illustration
is given through the case of a three-level  system coupled to the heat-bath both in the almost Markovian and strongly non-Markovian  regimes.

\section*{Acknowledgment}
This work was supported by RFBR and DFG. The IN2P3(France)-JINR(Dubna) Cooperation
Program is gratefully acknowledged.

\end{document}